\documentclass{an}
\usepackage{graphicx}
\usepackage{times}
\usepackage{enumerate}
\usepackage{natbib}
\bibpunct{(}{)}{;}{a}{}{,}
\begin{document}

\Pagespan{789}{}
\Yearpublication{2006}%
\Yearsubmission{2005}%
\Month{11}%
\Volume{999}%
\Issue{88}%

\title{Spectra of faint sources in crowded fields with FRODOSpec on the \\
Liverpool Robotic Telescope}
\author{V.N. Shalyapin\inst{1,2}
\and  
L.J. Goicoechea\inst{1}\fnmsep\thanks{Corresponding author: 
\email{goicol@unican.es}\newline}}

\titlerunning{Spectra of faint sources in crowded fields with FRODOSpec}
\authorrunning{V.N. Shalyapin \& L.J. Goicoechea}

\institute{
Departamento de F\'\i sica Moderna, Universidad de Cantabria, 
Avda. de Los Castros s/n, 39005 Santander, Spain
\and 
Institute for Radiophysics and Electronics, National Academy of 
Sciences of Ukraine, 12 Proskura St., 61085 Kharkov, Ukraine}



\abstract{We check the performance of the FRODOSpec integral-field spectrograph for 
observations of faint sources in crowded fields. Although the standard processing pipeline L2 
yields too noisy fibre spectra, we present a new processing software (L2LENS) that gives rise 
to accurate spectra for the two images of the gravitationally lensed quasar Q0957+561. Among 
other things, this L2LENS reduction tool accounts for the presence of cosmic-ray events, 
scattered-light backgrounds, blended sources, and chromatic source displacements due to 
differential atmospheric refraction. Our non-standard reduction of Q0957+561 data shows the 
ability of FRODOSpec to provide useful information on a wide variety of targets, and thus, 
the big potential of integral-field spectrographs on current and future robotic telescopes.}

\maketitle

\section{Introduction}
\label{sec:1}

The Fibre-fed RObotic Dual-beam Optical Spectrograph \citep[FRODOSpec;][]{morales04} is the 
multi-purpose spectrograph on the Liverpool Robotic Telescope \citep{steele04}. FRODOSpec has 
two independent arms, allowing simultaneous spectroscopy at blue and red wavelengths. It is 
also equipped with an integral field unit, which consists of 12$\times$12 square lenslets 
(microlenses) each 0\farcs83 on sky, bonded to 144 optical fibres and covering a field of view 
of about 10\arcsec$\times$10\arcsec. After an observation session, the data processing 
pipeline L2 \citep{barnsley12b} automatically extracts a wavelength-calibrated spectrum for 
each fibre. Later, the user can combine some of these raw (sky-unsubtracted and 
flux-uncalibrated) fibre spectra, subtract the background sky level, apply a flux calibration 
to the sky-subtracted data, and so on. An automatic sky subtraction is also possible when L2 
successfully identifies sky-only fibres. 

FRODOSpec was designed mainly to study bright point-like sources \citep{morales04}, and the 
L2 outputs for point-like sources with $V <$ 12 mag are leading to high quality spectra 
\citep{camero12,casares12,barnsley13,ribeiro13}. However, L2 has been developed to produce 
quick look data instead of optimal spectral results. Additionally, the ability of this 
spectrograph to render accurate spectra of fainter and/or blended sources has not been 
explored in detail so far, and only \citet{nugent11} have presented a useful spectrum of a 
point-like source with $V$ = 15 mag (SN 2011fe). In this paper, we focus on a typical 
observation session of our pilot project to follow-up the spectrophotometric variability of 
\object{Q0957+561} \citep{walsh79} with the Liverpool Robotic Telescope. \object{Q0957+561} is 
a gravitational lens system consisting of a lensed quasar with two relatively faint images ($V 
\sim$ 17 mag) and a lensing elliptical galaxy. 
         
In Sect.~\ref{sec:2} we describe the relevant properties of the science target, as well as 
the standard processing pipeline L2 and its outputs for the typical observation session. This 
standard pipeline does not accurately extract the raw spectrum for each fibre for the 
observation session. In addition, the spectrophotometry of gravitational lens systems 
requires some steps that are not incorporated into L2. Thus, in Sect.~\ref{sec:3} we 
introduce a new processing method (L2LENS), which is designed to obtain flux-calibrated 
spectra of faint sources in crowded fields. In Sect.~\ref{sec:4} we obtain the L2LENS 
spectrum for each source in \object{Q0957+561}. The conclusions are presented in 
Sect.~\ref{sec:5}.  

\section{FRODOSpec data of Q0957+561 and L2}
\label{sec:2}

\object{Q0957+561} (our science target) consists of two quasar images, A and B, separated by 
6\farcs1 with identical redshifts ($z_{\rm{Q}}$ = 1.41), and a lensing elliptical galaxy 
($z_{\rm{L}}$ = 0.36) placed between A and B \citep{walsh79,young80}. The G (lensing) galaxy 
is only $\sim$ 1\arcsec\ apart from B. Accurate positions of the B image and G relative to the 
A image, and the optical structure of G, i.e., its de Vaucouleurs profile, are known from 
Hubble Space Telescope (HST) observations \citep{bernstein97,keeton98,kochanek13}. While both 
quasar images have roughly similar brightness in optical bands, the galaxy becomes 
increasingly bright with increasing wavelength, having comparable brightness to those of A and 
B at the reddest optical wavelengths.  

\begin{figure}
\includegraphics[width=\linewidth]{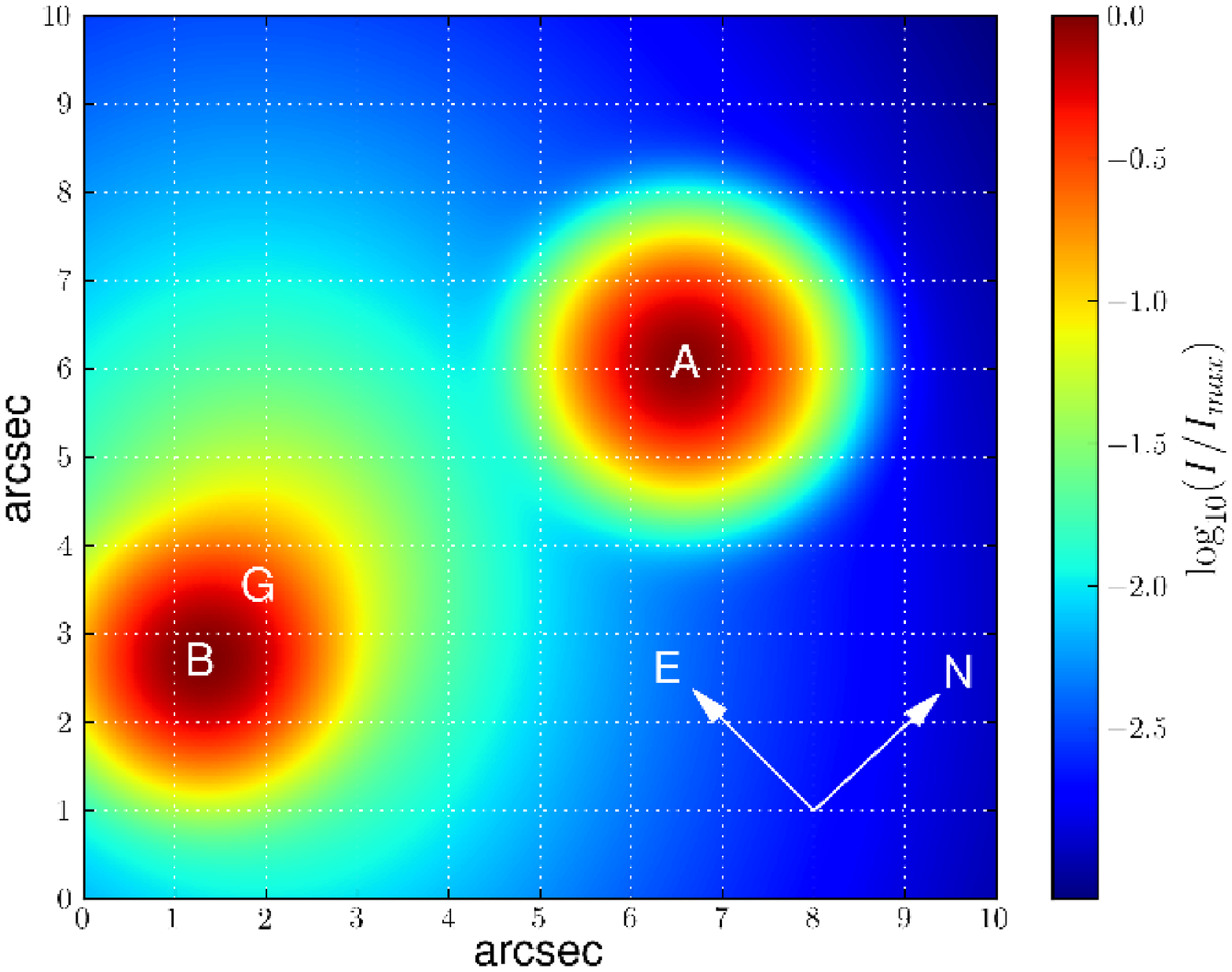}
\includegraphics[width=\linewidth]{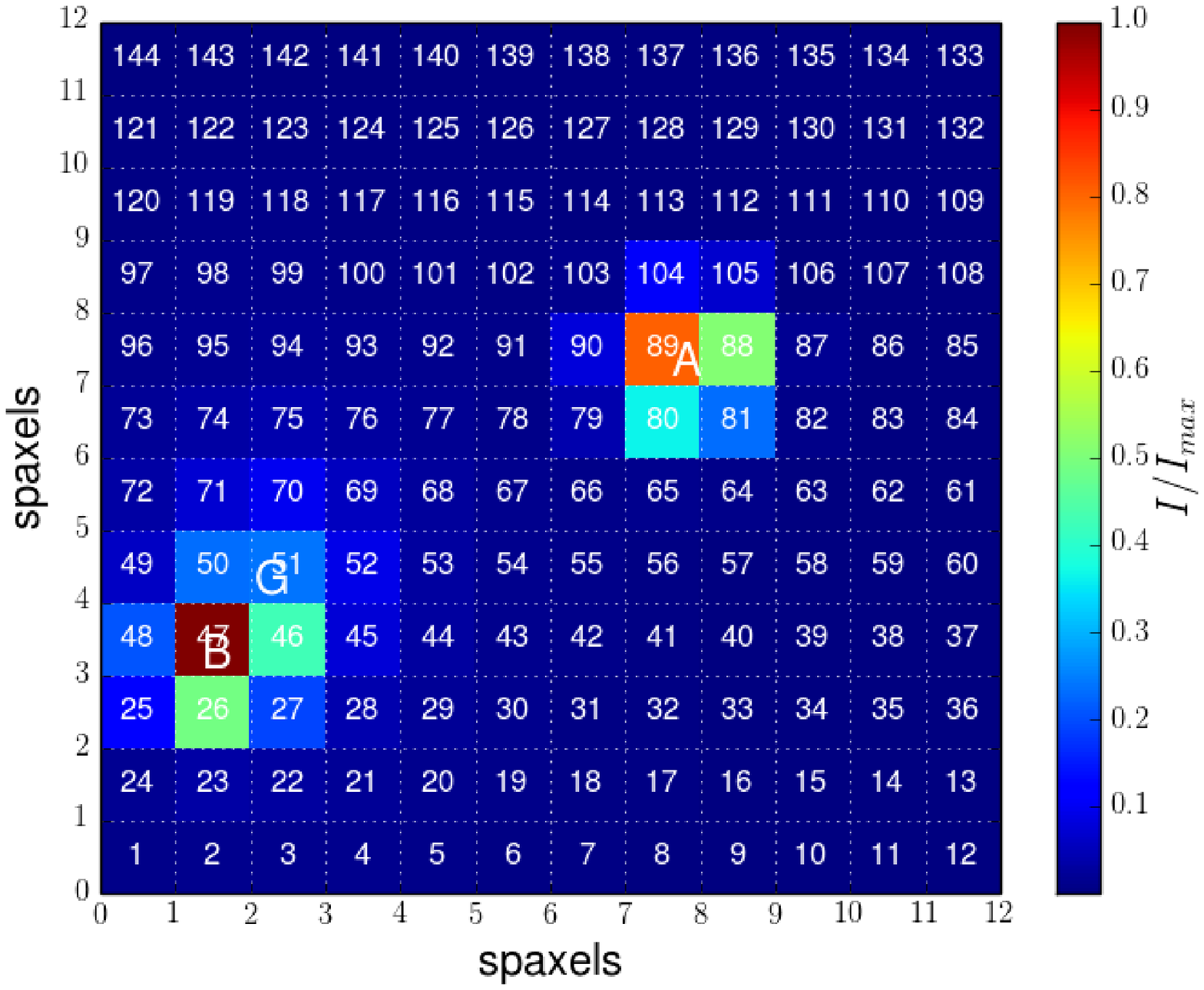}
\caption{Model of Q0957+561 on the integral field unit of FRODOSpec. Top: the three sources 
(A, B and G) convolved with a 2D-Gaussian PSF (FWHM = 1\farcs2). We use astro-photometric 
data of the system at the reddest optical wavelengths and a simple model for the atmospheric 
seeing. Bottom: integrated brightness within each lenslet. The whole array contains 144 
(12$\times$12) lenslets coupled to fibres. Here, "spaxels" means spatial pixels.}
\label{Fig1}
\end{figure}

Due to extended emission in G and atmospheric blurring, there is some level of 
cross-contamination between the three sources A, B and G. In Fig.~\ref{Fig1}, we show a model 
of \object{Q0957+561} on the integral field unit of FRODOSpec (see Sect.~\ref{sec:1}). This 
model (simulation) considers all astrometric and photometric properties of the system at the 
red end of the optical spectrum (A, B and G are of equal brightness), as well as the 
convolution between source profiles and a two-dimensional Point Spread Function (PSF) having a 
Gaussian shape with FWHM  = 1\farcs2. Mutual contamination between blurred point-like and 
extended sources is evident in the top panel of Fig.~\ref{Fig1}, which displays the brightness 
distribution on a logarithmic scale. The finite size of the 144 square lenslets also brings a 
pixelation effect over the field of view. The bottom panel of Fig.~\ref{Fig1} shows the 
integrated brightness within each spatial pixel. After rearranging the fibre bundle in a 
zigzag pattern (see labels in the bottom panel of Fig.~\ref{Fig1}), the spectrograph behaves 
as a linear pseudo-slit.

During the \object{Q0957+561} monitoring program, we have obtained data in many observation 
sessions using the low-resolution configuration. This configuration enables users to get 
spectra with wavelength ranges (resolving powers) of 3900$-$5700 \AA\ (2600) and 5800$-$9400 
\AA\ (2200) for the blue and red arms. Here, in order to demonstrate the problems of the 
standard processing pipeline L2 with our science exposures, and the need for a new reduction 
scheme suitable for blended faint sources in crowded fields, we concentrate on the session on 
2011 March 1. Details of the observation log (blue arm) for this night are presented in 
Table~\ref{table1}. The first letter of the filenames ('b') refers to the blue arm. Files 
from the red arm start with the letter 'r'. The second letter of the filenames denotes the 
target, i.e., 'w', 'e' and 'a' for Tungsten continuum lamp (hereafter W lamp; these exposures 
are used as flats for tracing fibres), sky target (hereafter q0957 $\equiv$ 
\object{Q0957+561} or feige34 $\equiv$ \object{Feige 34}; feige34 is the calibration star) 
and Xenon arc lamp (hereafter Xe lamp; exposures for wavelength calibrations), respectively. 
All blue-arm and red-arm data files may be downloaded from the Gravitational LENses and DArk 
MAtter (GLENDAMA) website\footnote{http://grupos.unican.es/glendama/LQLM\_tools.htm} (see 
Appendix A).

\begin{table}
\caption{Details of the observation log (blue arm) on 2011 March 1}
\begin{tabular}{lccl}
\hline
Spectrum       &  UT       &  Exposure     &  Filename (FITS format)       \\
               & (hh:mm)   &  (s)          &                               \\
\hline
W lamp 	   &  19:22    &    60         &  b\_w\_20110301\_2\_1\_1\_1   \\
q0957          &  21:07    &  2700         &  b\_e\_20110301\_7\_1\_1\_2   \\
Xe lamp        &  21:52    &    60         &  b\_a\_20110301\_8\_1\_1\_1   \\
feige34        &  21:57    &   100         &  b\_e\_20110301\_9\_1\_1\_2   \\
Xe lamp        &  21:59    &    60         &  b\_a\_20110301\_10\_1\_1\_1  \\
\hline
\\
\end{tabular}
Notes: q0957 was observed at an air mass of 1.43 and a Moon fraction of 0.08. At the start of 
observations, the estimated seeing was FWHM = 1\farcs14. The star (feige34) was observed in 
similar conditions.  
\label{table1}
\end{table}

\begin{figure}
\includegraphics[width=\linewidth]{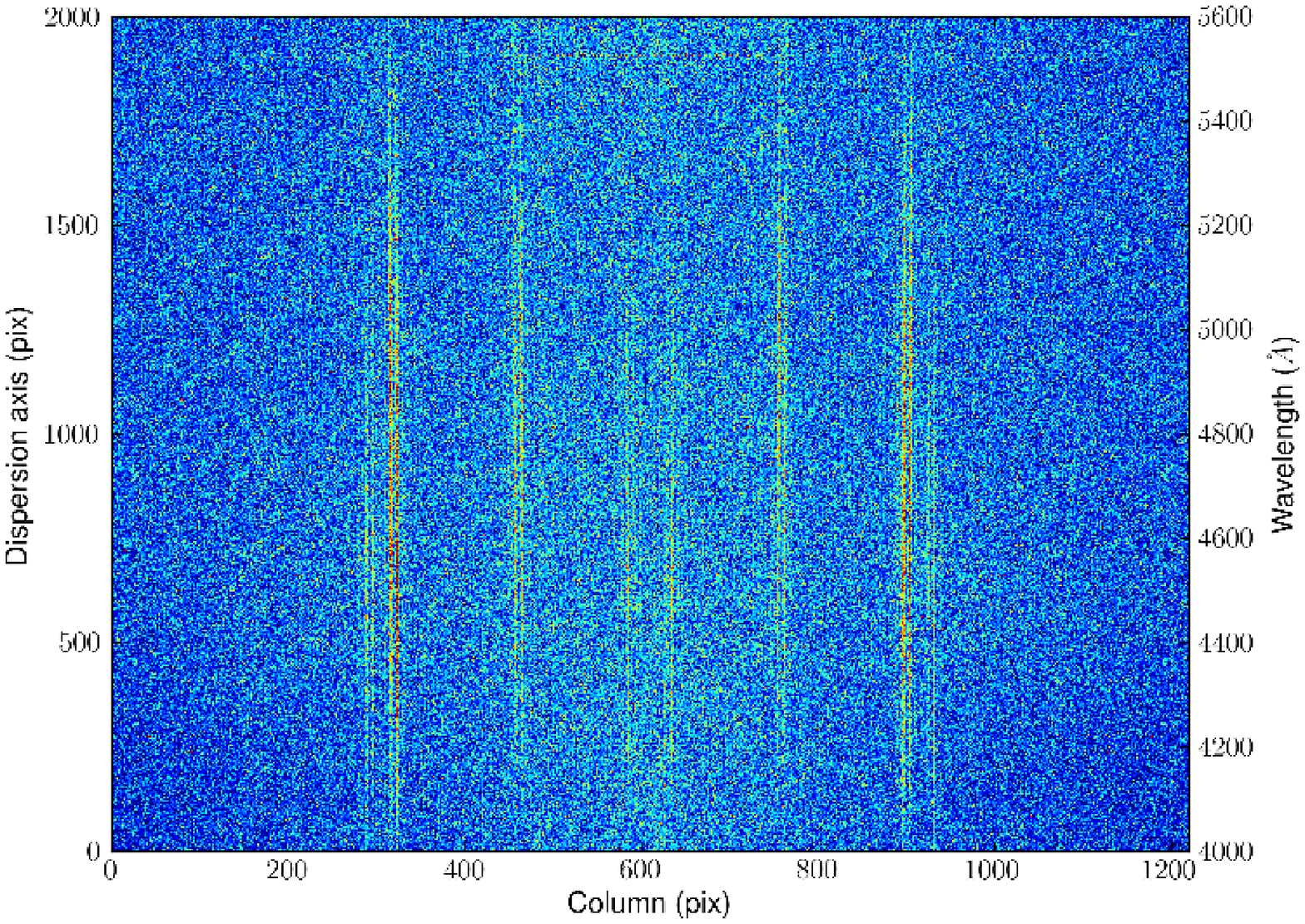}
\includegraphics[width=\linewidth]{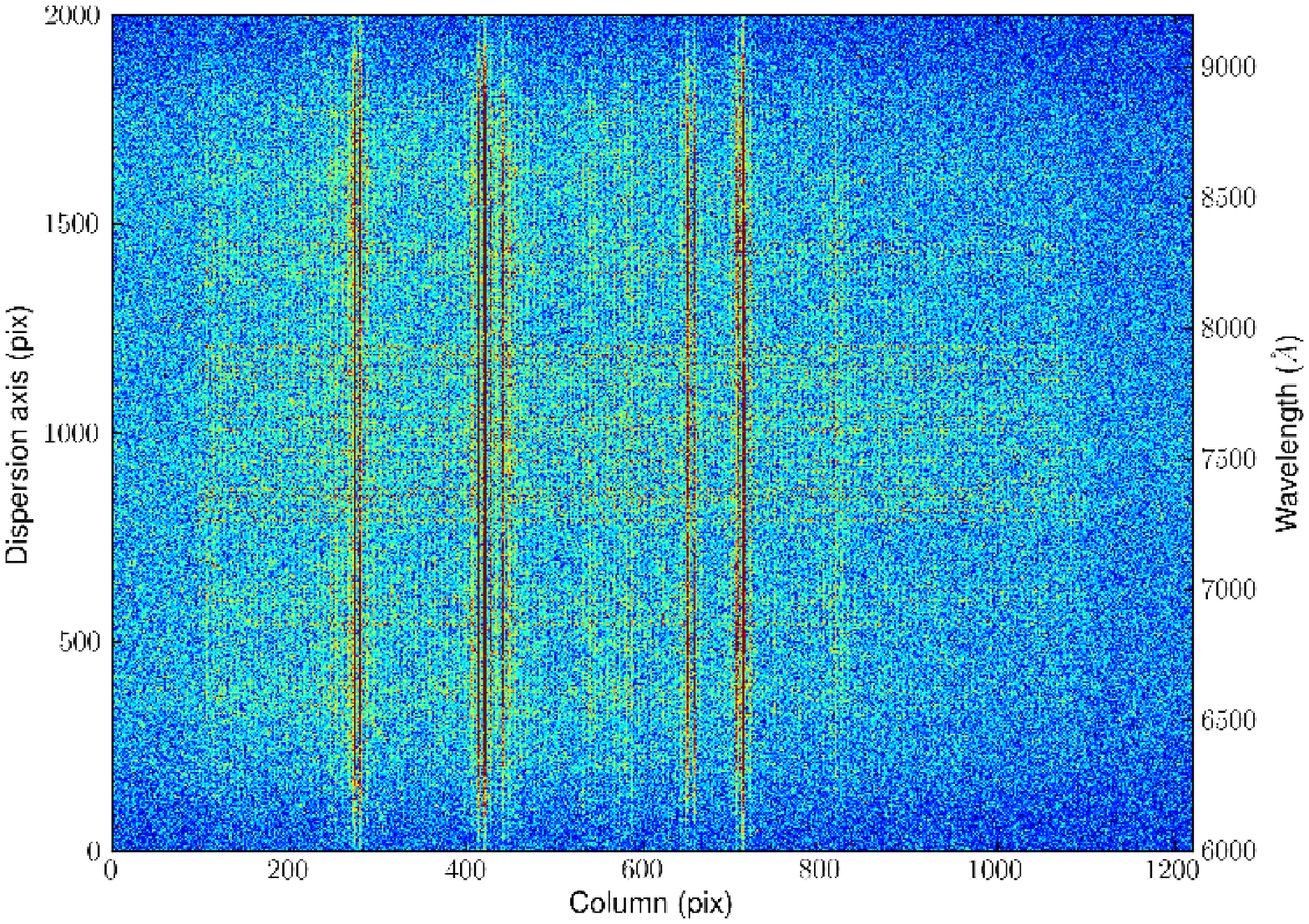}
\caption{Outputs of the L1 pipeline for q0957. Top: blue spectral arm. Bottom: red spectral 
arm. See main text for details.}
\label{Fig2}
\end{figure}

Data taken by FRODOSpec are reduced by two pipelines. The first pipeline, known as L1, is a 
CCD pre-processing task. L1 performs bias subtraction, overscan trimming and CCD flat 
fielding. For a sky target in a given arm, the output from L1 is saved as the zero extension 
in an eight part multi-extension FITS file, e.g., b\_e\_20110301\_7\_1\_1\_2.fits (see 
Table~\ref{table1}) stores q0957 data in eight extensions [0$-$7], where [0] contains the L1 
output. The L1 outputs for q0957 in both spectral arms are shown in Fig.~\ref{Fig2}. In this 
figure, the pseudo-slit is oriented along rows (cross-dispersion axis), whereas columns 
correspond to the dispersion axis. The 144 fibres are located between columns 70 and 1080 in 
the blue arm, with an inter-fibre space of about 7 pixels (see the top panel of 
Fig.~\ref{Fig2}). In the red arm, the fibres are displaced 20 pixels to the right (see the 
bottom panel of Fig.~\ref{Fig2}). 

\begin{figure}
\includegraphics[width=\linewidth]{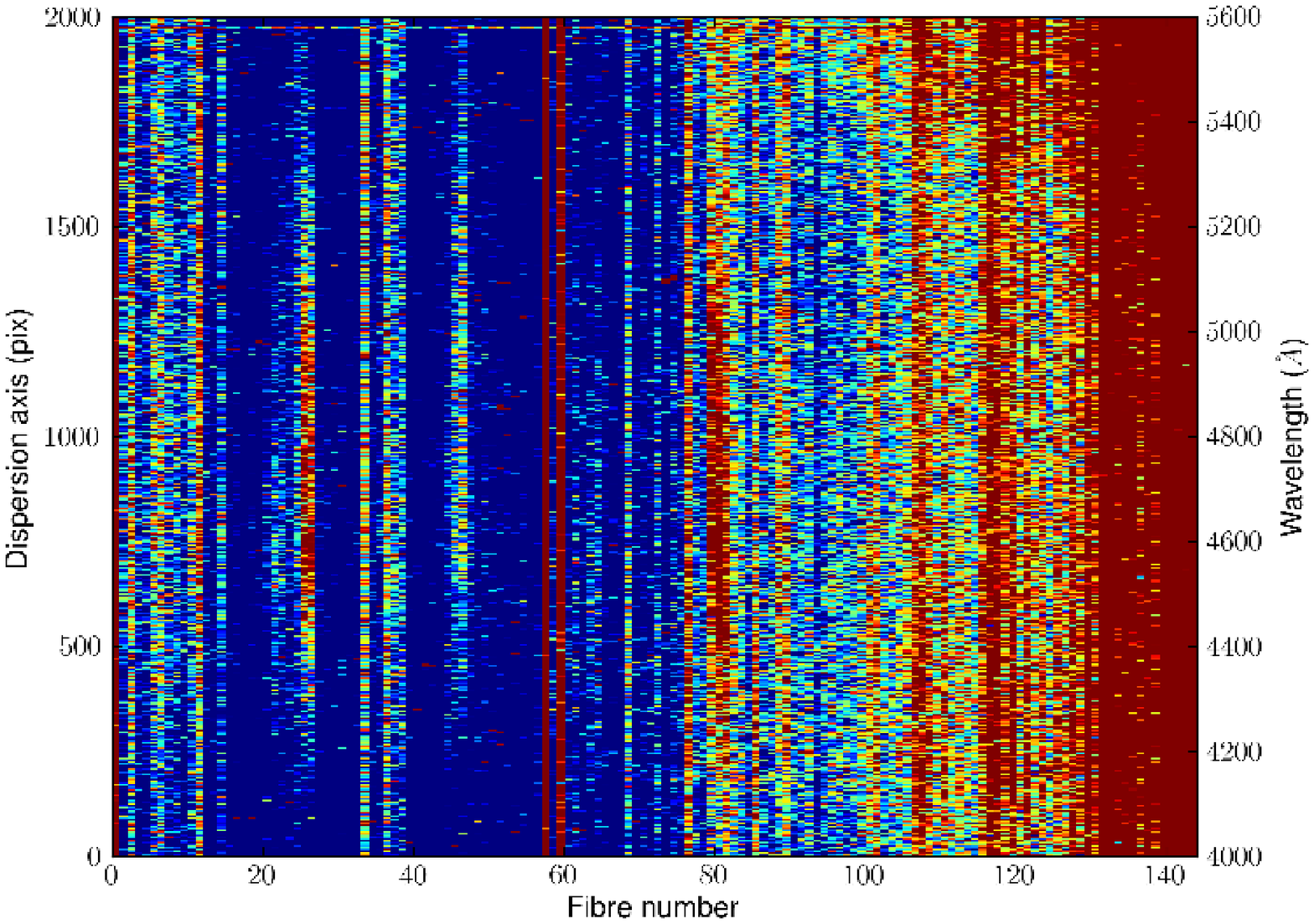}
\includegraphics[width=\linewidth]{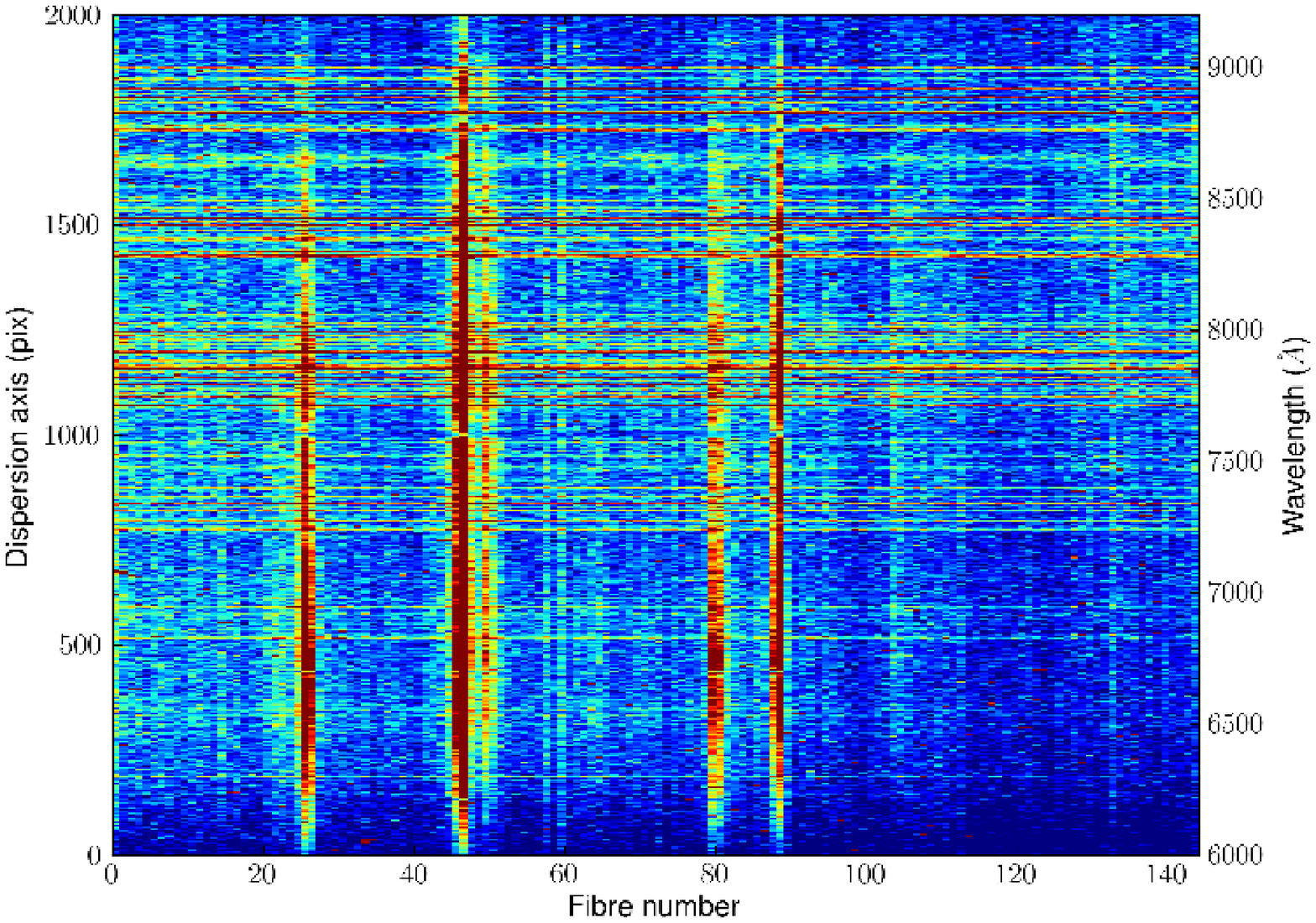}
\caption{Outputs of the L2 pipeline for q0957. Top: raw spectra (one per fibre) in the blue 
arm. Bottom: raw spectra in the red arm. See main text for details.}
\label{Fig3}
\end{figure}

The second pipeline, known as L2, performs the tasks unique to Integral Field Spectroscopy 
(IFS) reduction, and requires three frames to proceed: a sky target frame (q0957 or 
feige34), and W and Xe lamp exposures. This standard processing pipeline includes the 
following main steps \citep{barnsley12b}:
\begin{enumerate}[(i)]
  \item to find and trace the position of each fibre along the dispersion axis (fibre 
tramline map generation from the W lamp exposure and polynomial fits);
  \item to extract the instrumental flux in a 5-pixel aperture around the position of each 
fibre along the dispersion axis (standard aperture flux extraction in the W and Xe lamp 
exposures, and the sky target frame using the fibre tramline map);
  \item wavelength calibration on a fibre-to-fibre basis (using the Xe lamp spectrum for each 
fibre);
  \item fibre transmission correction (using the W lamp spectrum for each fibre, 
fibre-to-fibre throughput differences in the sky target spectra are properly corrected);
  \item single wavelength solution for all fibres (rebinning the flux along the dispersion 
axis for each sky target spectrum). 
\end{enumerate}
At this stage, the outputs from L2 are saved as the extensions [1] and [2] in the 
corresponding multi-extension FITS file. For example, b\_e\_20110301\_7\_1\_1\_2.fits[1] 
contains the 144 wavelength-calibrated spectra for q0957 in the blue arm, while 
b\_e\_20110301\_7\_1\_1\_2.fits[2] is a spectral data cube. This data cube gives the 2D flux 
in the 12$\times$12 fibre array at each wavelength pixel. We remark that the [1$-$2] 
extensions comprise sky-unsubtracted and flux-uncalibrated spectra. 

Unfortunately, the results for q0957 are not of sufficient quality, and these can be 
significantly improved (see Sect.~\ref{sec:3}). The outputs of the L1 pipeline for q0957 
(Fig.~\ref{Fig2}) clearly show several bright vertical lines associated with the signal from 
the lens system, whereas the subsequent L2 outputs (one raw spectrum for each fibre on each 
arm) are rather confusing (Fig.~\ref{Fig3}). In the bottom panel of Fig.~\ref{Fig3}, the red 
light of the science target is gathered near the fibres 26, 47 and 50 (B image plus galaxy), 
and 80 and 90 (A image), but in the top panel of Fig.~\ref{Fig3}, is it hard to identify the 
blue light from the quasar images. 

\section{Scattered light subtraction and other refinements}
\label{sec:3}

\begin{figure}
\includegraphics[width=\linewidth]{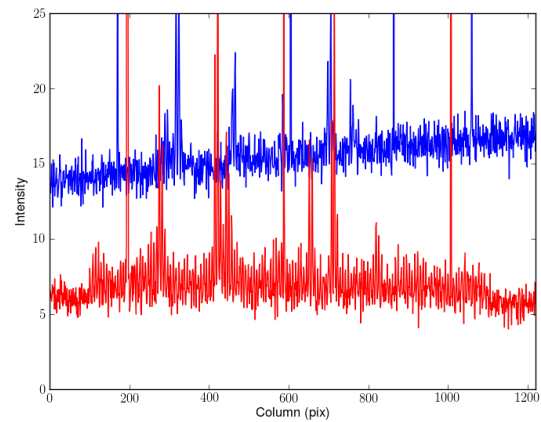}
\caption{Spatial distributions of light at the central rows of the dispersion axis. Top: blue 
spectral arm. Bottom: red spectral arm. See main text for details.}
\label{Fig4}
\end{figure}

It is worth examining more closely the frames in Fig.~\ref{Fig2}. A detailed look at these L1 
outputs reveals two important features. Firstly, there are numerous cosmic rays. As noted by 
\citet{barnsley12b}, automated removal of cosmic rays from spectrographic data is too 
unreliable. However, manual removal of cosmic rays with visual inspection of results works 
quite well. In particular, we have found that the spectral version of the L.A.Cosmic 
algorithm \citep{vandokkum01} provides good results \citep[see also the PyCosmic algorithm 
by][]{husemann12}. The second feature is more critical. Long exposures (see the exposure 
time for q0957 in Table~\ref{table1}) lead to high background levels, as shown in 
Fig.~\ref{Fig4}. This figure traces the flux curves along the cross-dispersion axis at the 
central rows of each arm, or more exactly, the spatial distributions of light, averaging over 
the rows 991$-$1010 in the two spectral arms. Apart from sharp spikes at several columns 
caused by cosmic rays, there are significant background levels in the fibres between the 
columns 70 and 1100, the inter-fibre regions (out of the fibre wings), and the fibre-free 
area encompassing columns 1$-$70 and 1100$-$1220. Hence, we are dealing with scattered 
(stray) light backgrounds \citep[e.g.,][]{sanchez06a}, which have not been subtracted by the 
standard pipeline L2. These 15-count (blue arm) and 7-count (red arm) average background 
levels (see Fig.~\ref{Fig4}) can be compared with useful signals in both arms to roughly 
estimate signal-to-background ratios of about 0.5 (blue arm) and 2 (red arm). In the blue 
arm, the spectral information is substantially degraded in the presence of large amounts of 
non-uniformly scattered light. The vertical artefacts apparent in the top panel of 
Fig.~\ref{Fig3} are due to backgrounds in fibres and their (de)magnifications through the 
step (iv) in L2.  

We have developed procedures for cleaning the L1 outputs of cosmic rays and scattered light. 
These procedures are part of a new processing scheme called L2LENS, in which the extraction 
of fibre spectra is mainly based on the SPECRED package of the Image Reduction and Analysis 
Facility (IRAF)\footnote{IRAF is distributed by the National Optical Astronomy Observatory, 
which is operated by the Association of Universities for Research in Astronomy (AURA) under 
cooperative agreement with the National Science Foundation. This software is available at 
http://iraf.noao.edu/}. Apart from IRAF commands, we also use Python\footnote{Python was 
created in the early 1990s by Guido van Rossum. Since 2001, the Python Software Foundation 
promotes, protects, and advances the Python programming language, as well as supports and 
facilitates the growth of an international community of Python programmers. This software is 
available at http://www.python.org/} scripts to perform reduction tasks. The L2LENS scripts
(see Appendix A) are freely available at the GLENDAMA website$^1$. With respect to the fibre 
spectrum extraction, the main differences between the 
standard L2 \citep[][and Sect.~\ref{sec:2}]{barnsley12a,barnsley12b} and the new L2LENS are:
\begin{enumerate}[(i)]
\item cosmic rays are removed by the spectral version of 
L.A.Cosmic\footnote{http://www.astro.yale.edu/dokkum/lacosmic/lacos\_spec.cl};
\item scattered light subtraction is carried out by means of the IRAF SPECRED/APSCATTER task, 
which fits a two-dimensional polynomial function to the inter-fibre background\footnote{For 
each pair of adjacent fibres whose centres are separated by about 7 pixels, the inter-fibre 
region is basically defined as the pixel located amid the two fibres}. To perform this fit, 
fifth-order spline functions along the cross-dispersion (columns in Fig.~\ref{Fig2}) and 
dispersion (rows in Fig.~\ref{Fig2}) axes are used; 
\item alternative for the step (i) in L2: to trace the fibre positions near the blue end of 
the red arm, the polynomial order is set to APALL.T\_ORDER = 6;
\item alternative for the step (ii) in L2: fibre flux extraction is made by using a 4-pixel 
aperture rather than the standard one (5 pixels). This procedure tries to avoid the noise in 
the edges of the standard aperture\footnote{The optimal flux extraction \citep{horne86} does 
not lead to a substantial improvement. However, there are more refined techniques of 
extraction using a cross-dispersion profile fitting \citep[e.g.,][]{sandin10}};  
\item wavelength calibration is done in two stages. Firstly, a dispersion solution is found 
for each fibre independently. A cubic polynomial smoothing is then used to generate a global
calibration from the set of 144 initial solutions. In the red arm, the smoothing procedure 
noticeably reduces fringe effects; 
\item fibre-to-fibre and wavelength-to-wavelength throughput differences are corrected. Using
the W lamp spectrum for each fibre, an average spectrum is calculated, and then smoothed by a
20-point filter. Individual W lamp spectra are divided by this smoothed (average) spectrum to
get correction coefficients for each fibre and wavelength pixel.
\end{enumerate}

\begin{figure}
\includegraphics[width=\linewidth]{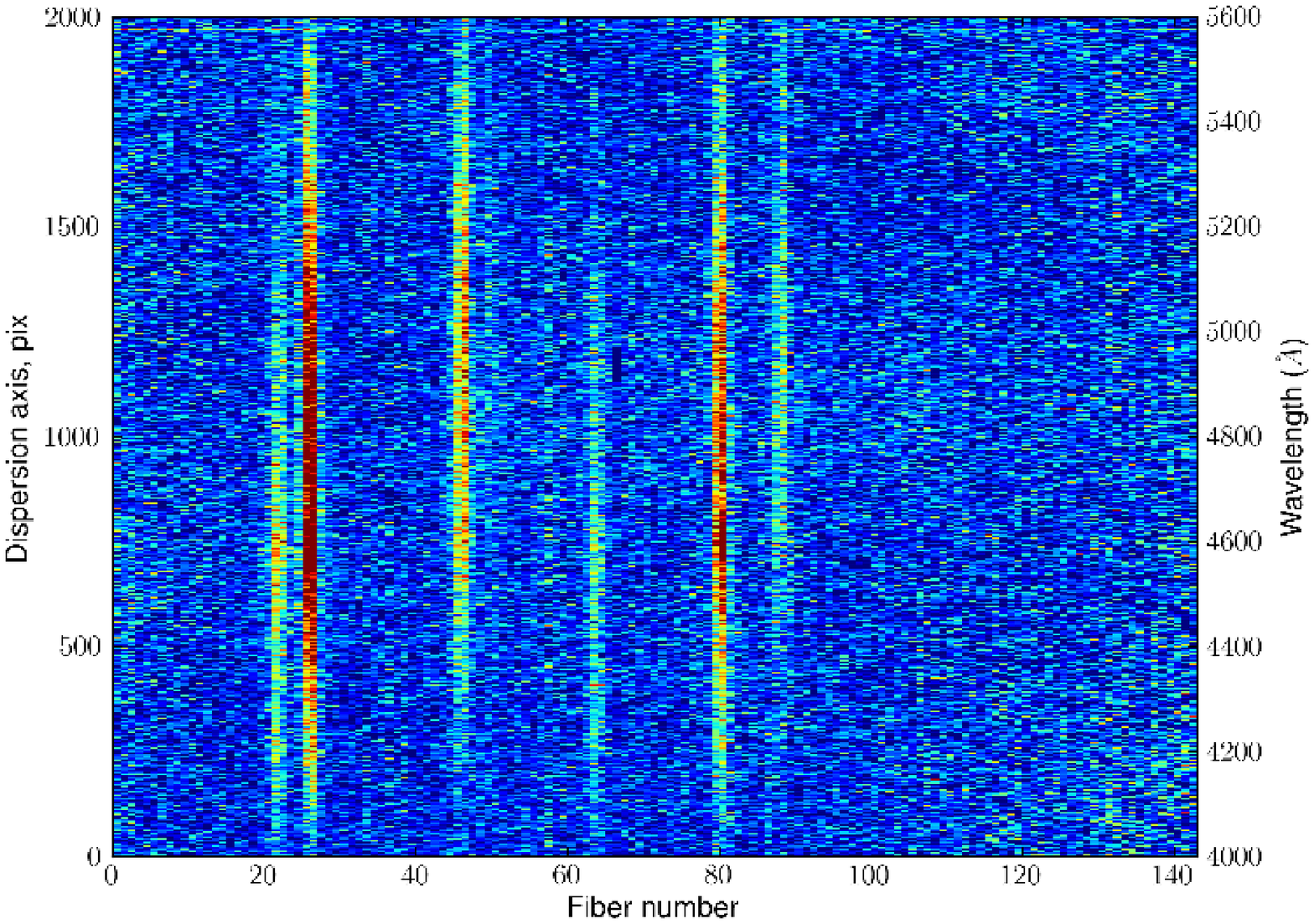}
\includegraphics[width=\linewidth]{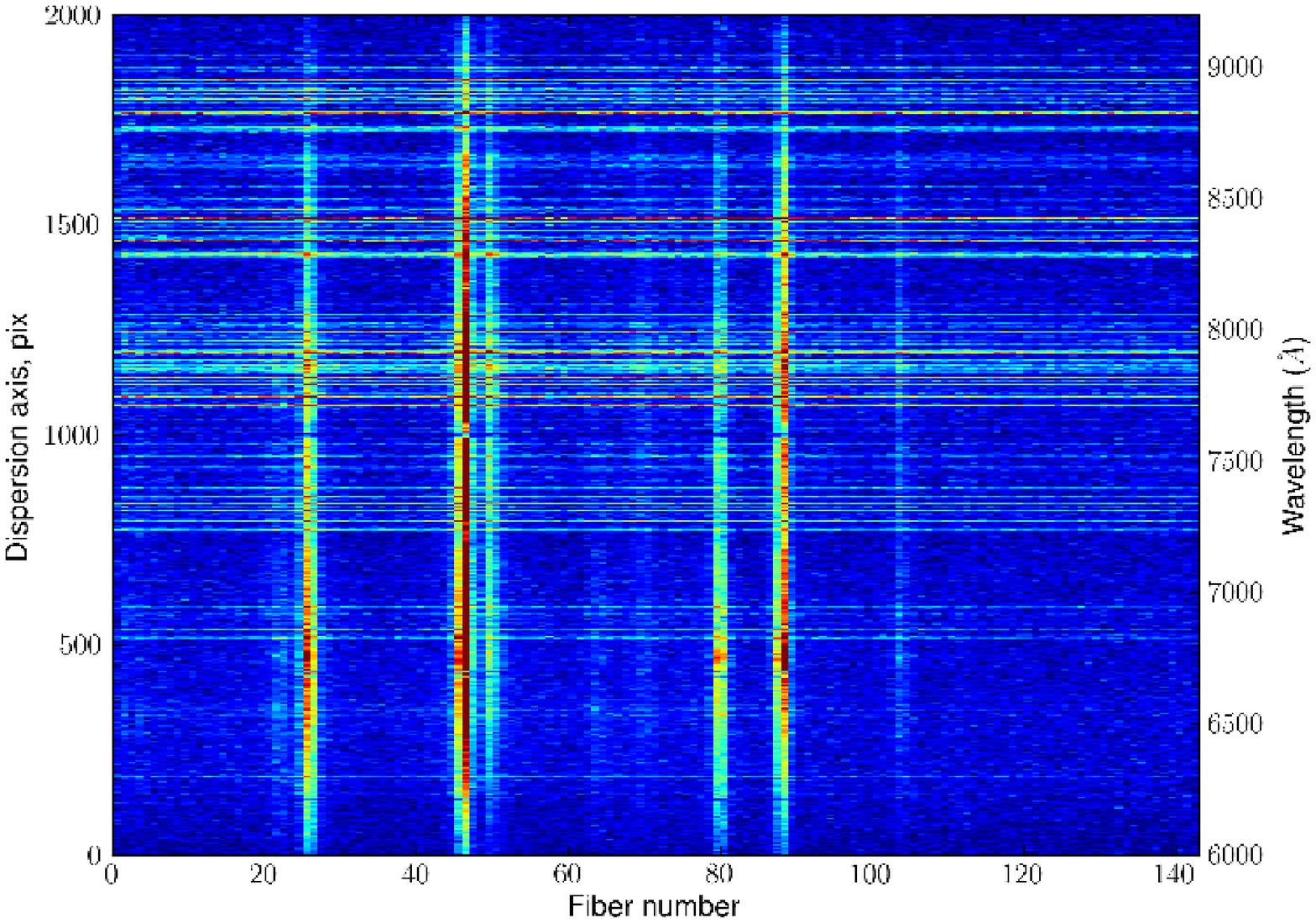}
\caption{Intermediate outputs of the L2LENS software for q0957. Top: raw spectra (one per 
fibre) in the blue arm. Bottom: raw spectra in the red arm. See main text for details.}
\label{Fig5}
\end{figure}

\begin{figure}
\includegraphics[width=\linewidth]{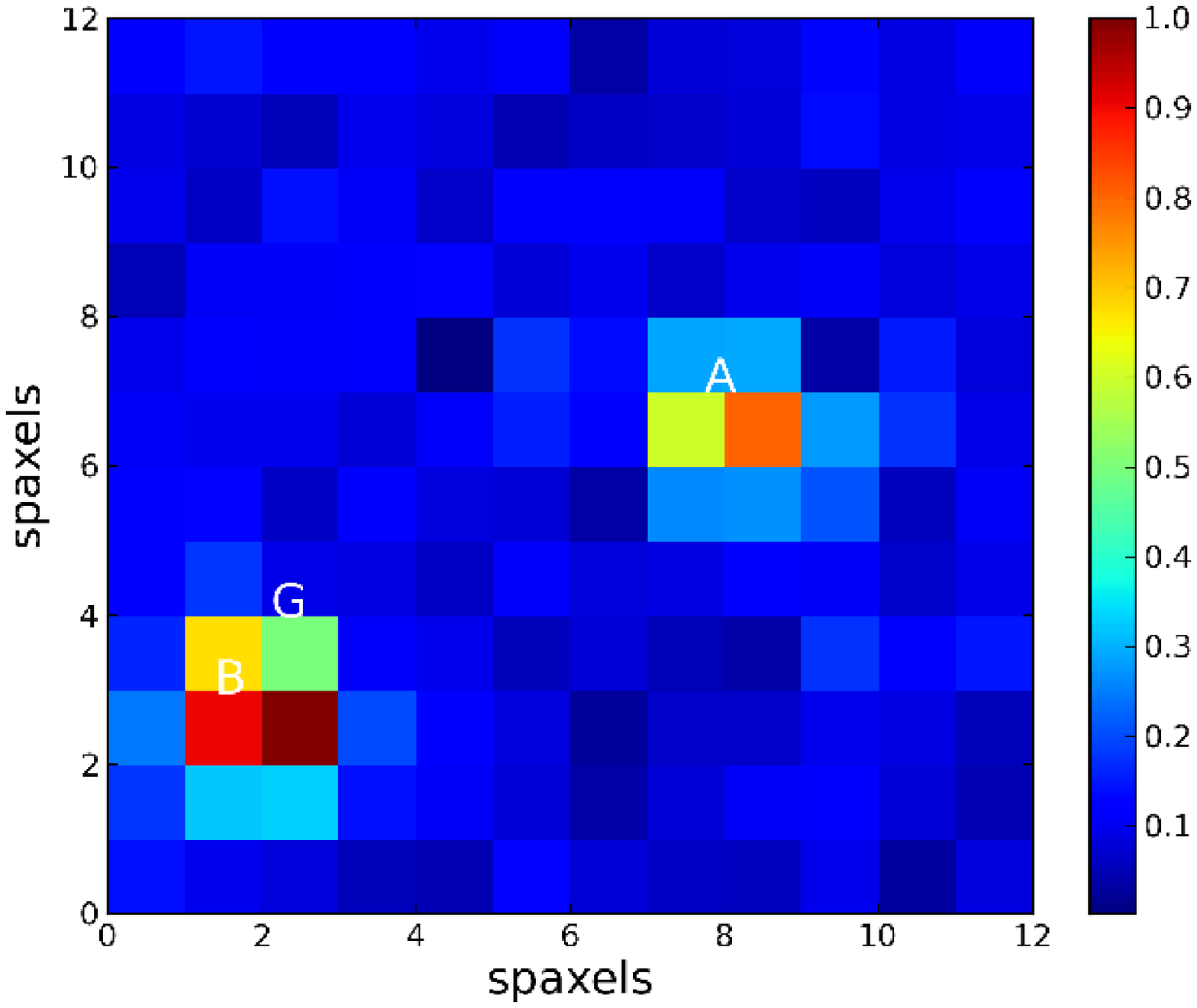}
\includegraphics[width=\linewidth]{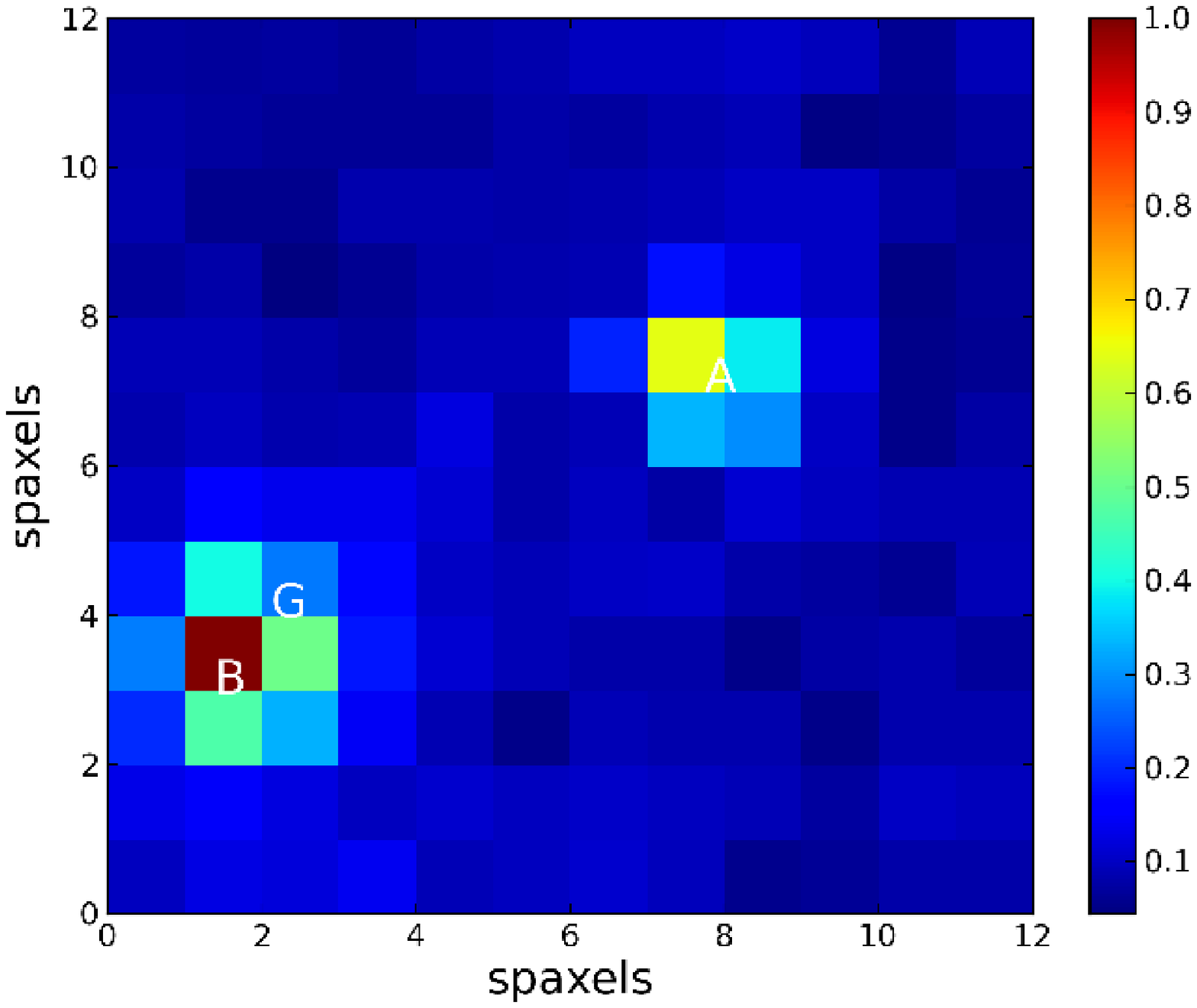}
\caption{Monochromatic frames of q0957 from the processing scheme L2LENS. Top: $\lambda$ = 
4800 \AA\ (blue arm). Bottom: $\lambda$ = 7600 \AA\ (red arm).}
\label{Fig6}
\end{figure}

At this intermediate stage of L2LENS, its outputs are shown in Fig.~\ref{Fig5}. These must be 
compared with the L2 products in Fig.~\ref{Fig3}. There is a substantial improvement in the 
blue arm (see the top panels of Fig.~\ref{Fig3} and Fig.~\ref{Fig5}), while changes in the 
red arm are not so evident. Although a visual inspection of the bottom panels of 
Fig.~\ref{Fig3} and Fig.~\ref{Fig5} does not reveal clear differences between both outputs, a
quantitative analysis also indicates an appreciable improvement in the red arm with the use 
of L2LENS. In addition, we can construct spectral data cubes (see Sect.~\ref{sec:2}). Each 
data cube consists of 12$\times$12 individual spectra (one per fibre) with 2001 wavelength 
pixels. The wavelength intervals (dispersions) are 4000$-$5600 \AA\ (0.8 \AA/pixel) and 
6000$-$9200 \AA\ (1.6 \AA/pixel) for the blue and red arms. Here, we do not consider the 
spectral edges in the wavelength ranges 3900$-$4000, 5600$-$5700, 5800$-$6000 and 9200$-$9400 
\AA\ (see the third paragraph in Sect.~\ref{sec:2}) because they include very noisy, unusable 
data. Fig.~\ref{Fig6} displays the monochromatic frames of q0957 at the central wavelengths 
of both arms, i.e., 4800 \AA\ (top panel) and 7600 \AA\ (bottom panel). 

\section{Flux-calibrated spectra of the lens system}
\label{sec:4}

When doing photometry on crowded fields, such as lens systems, aperture photometry does not 
yield reliable results. It is better to use a PSF fitting method 
\citep{wisotzki03,becker04,roth04}. An additional complication arises as a result of the 
differential atmospheric refraction, which causes source position variations with wavelength 
\citep{filippenko82}. For example, \citet{arribas99} introduced an interpolation procedure to 
correct these effects in IFS. Our L2LENS software incorporates a PSF fitting task that takes 
differential atmospheric refraction effects into account.

\subsection{PSF fitting method}
\label{sec:4.1}

In order to obtain accurate fluxes for the two quasar images of the lens system q0957, we use 
a method similar to that presented in \citet{wisotzki03}. The Wisotzki et al.'s procedure was 
successfully applied to IFS data of the quadruply lensed quasar HE 0435$-$1223. At each 
wavelength (monochromatic frame), \citet{wisotzki03} decomposed this lens system into four 
point-like sources (quasar images) convolved with analytical 2D-Gaussian PSFs, plus a 
spatially constant background. In other words, each quasar image in each monochromatic frame 
was initially characterised by six free parameters: centroid, FWHM along major and minor 
axis, position angle and amplitude. However, Wisotzki et al. used some priors to reduce the 
large number of initial free parameters, and thus, to simplify the $\chi^2$ minimization and
accurately determine the quasar fluxes. Firstly, the FWHM values were assumed to be the same 
for all four Gaussians. In a second step, the centroids, FWHM values and position angles were 
replaced by polynomial functions of $\lambda$, so only the four amplitudes were finally 
fitted. 

\begin{figure}
\includegraphics[width=\linewidth]{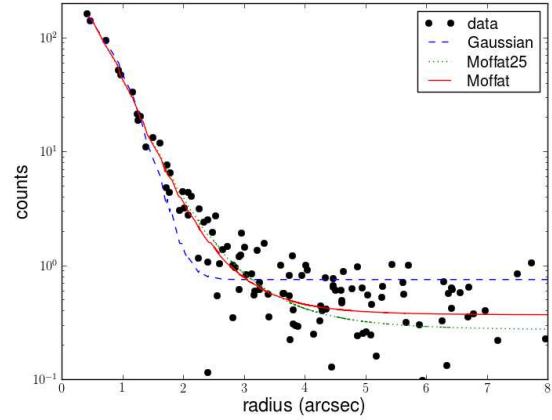}
\caption{Radial profile of the star feige34: observational data (circles), Gaussian fit 
(dashed line), $\beta$ = 2.5 Moffat fit (dotted line) and general Moffat fit (solid line).}
\label{Fig7}
\end{figure}

Our procedure is a variant of the Wisotzki et al.'s scheme. To model each monochromatic 
frame, we consider two point-like sources, as well as one extended (de Vaucouleurs profile) 
source and a uniform background. Hence, our model accounts for the presence of the lensing
galaxy. All three sources are convolved with the same PSF, and the first step is to decide on 
the most suitable shape of this convolution function. For example, \citet{sanchez06b} 
used an empirical PSF that was derived from a standard star. Unfortunately we find 
significant changes in the PSF with observing time, so the PSF of standard stars can not be 
incorporated into the model of the lens system. Whereas \citet{wisotzki03} used a 2D-Gaussian 
PSF, sometimes a Moffat profile is a better approach in IFS \citep{cairos12,kamann13}. Thus, 
in order to determine the optimal analytic shape of the PSF, we fit a red frame ($\lambda$ = 
7600 \AA) of the calibration star feige34 to different profiles. Fig.~\ref{Fig7} shows the 
observed radial profile of feige34 together with the fitted curves: Gaussian (dashed line), 
Moffat with $\beta$ (power-law index) = 2.5 (dotted line), and Moffat with an arbitrary 
$\beta$ (solid line). As can be seen in Fig.~\ref{Fig7}, the Gaussian curve underestimates 
the flux in the 1\farcs5$-$3\arcsec\ interval and overestimates the background level, but the 
Moffat curves work much better. The best fit corresponds to a Moffat profile with $\beta$ = 
2.9 (solid line), and we choose a $\beta$ = 3 Moffat distribution to model the PSF in the 
region of interest (lens system). 

Once the PSF shape is chosen, we also use an iterative fitting procedure. The positions of 
the B image and the lensing galaxy (G) relative to the A image, and the de Vaucouleurs 
profile of G are set to accurate values from HST observations 
\citep{bernstein97,keeton98,kochanek13}. Therefore, each monochromatic frame of q0957 is 
initially modelled by a ten-parameter distribution of light. These parameters are the 
centroid of A, the FWHM along major and minor axis, the position angle of the major axis, the 
orientation of the frame, the uncalibrated fluxes of A, B and G, and the background sky 
level. In a subsequent step, the first five parameters of each frame are set to the 
corresponding values of smoothly varying polynomial functions of $\lambda$, leaving only the
three uncalibrated fluxes and the sky level to be fitted (the orientation of the frame is set 
to the median of all orientations from the initial fits). For calibration purposes, we 
present the FRODOSpec spectral response function in Sect.~\ref{sec:4.2}. In Sect.~\ref{sec:4.3} 
we dive into detail on the PSF fitting task, and obtain the flux-calibrated spectra of A, B and G.   
 
\subsection{FRODOSpec spectral response function}
\label{sec:4.2}

\begin{figure}
\includegraphics[width=\linewidth]{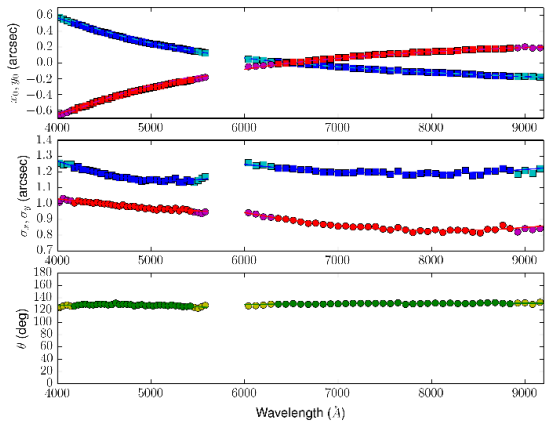}
\caption{PSF fitting results for 40 (blue arm) + 40 (red arm) spectral slices of feige34 
data. The five position-structure parameters of the Moffat distribution are fitted to smooth
polynomial functions, excluding the spectral edges in magenta, cyan and yellow (solid lines). 
In the top panel, we present the flux peak position ($x_{\rm{S}}$, $y_{\rm{S}}$) relative to 
that at $\lambda$ = 6500 \AA. We also show the expected values according to 
\citet{filippenko82} (dashed lines). See main text for details.}
\label{Fig8}
\end{figure}

PSF fitting is carried out on both the calibration star and the gravitational lens system. 
The calibration star model is defined by a single $\beta$ = 3 Moffat distribution (see 
Sect.~\ref{sec:4.1}) with a flux peak at ($x_{\rm{S}}$, $y_{\rm{S}}$), widths $\sigma_x$ and 
$\sigma_y$, a position angle of the major axis $\theta$, and a total flux $f_{\rm{S}}$. If 
($\hat{x}$, $\hat{y}$) is a coordinate system centered on the flux peak, and aligned with the 
major and minor axes, 
\begin{equation}
\hat{x} = (x-x_{\rm{S}})\cos \theta - (y-y_{\rm{S}})\sin \theta
\end{equation} 
\begin{equation}
\hat{y} = (x-x_{\rm{S}})\sin \theta + (y-y_{\rm{S}})\cos \theta  ,
\end{equation}
and the Moffat distribution with power-law index $\beta$ is given by
\begin{equation}
M(x,y) = f_{\rm{S}}\frac{\beta-1}{\pi \sigma_x \sigma_y} \left(1+r^2(x,y)\right)^{-\beta} ,
\label{Moffat}
\end{equation}
where
\begin{equation}
r(x,y) = \sqrt{\frac{\hat{x}^2}{\sigma_x^2}+\frac{\hat{y}^2}{\sigma_y^2}} .
\end{equation} 
For $\beta=3$, FWHM$_x$ = 1.02 $\sigma_x$ and FWHM$_y$ = 1.02 $\sigma_y$. The six parameters 
of $M(x,y)$ are calculated in an iterative manner. In the first iteration, both spectral data 
cubes (blue and red arms) with 12$\times$12$\times$2001 pixel$^3$ each are split into 40 
slices along the spectral axis. In each slice, 50 adjacent monochromatic frames are combined 
to increase the signal-to-noise ratio. In order to reduce sampling biases in $M(x,y)$ 
fitting, each of the 12$\times$12 spatial pixels (or spaxels) is divided into 10$\times$10 
equally-sized square subpixels, and the flux distribution in Eq.~\ref{Moffat} is evaluated on 
the 120$\times$120 subpixels of the fine mesh. A flux value for each fibre is then obtained 
by integrating over the associated spaxel (see Fig.~\ref{Fig1}). Our $\chi^2$ minimization 
procedure is based on the Levenberg-Marquardt algorithm. This calculates the six Moffat 
parameters ($x_{\rm{S}}$, $y_{\rm{S}}$, $\sigma_x$, $\sigma_y$, $\theta$, and $f_{\rm{S}}$) 
plus a uniform background. In the top panel of Fig.~\ref{Fig8}, chromatic changes in the flux 
peak position are due to differential atmospheric refraction, since the measurements of 
($x_{\rm{S}}$, $y_{\rm{S}}$) can be accounted for in terms of this atmospheric effect (dashed 
lines). In the middle panel of Fig.~\ref{Fig8}, we do not see the expected monotonic decrease 
of $\sigma_x$ with wavelength (blue and cyan squares). We note that the discontinuity does not 
appear in the $\sigma_x$ values from the lens system data (see below). Moreover, the anomaly 
is only observed some nights. At present, we are trying to find the origin of this 
time-dependent anomalous behaviour. 

\begin{figure}
\includegraphics[width=\linewidth]{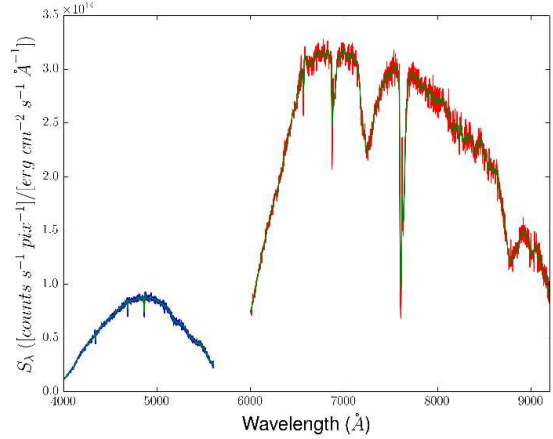}
\caption{FRODOSpec response function from feige34 data.}
\label{Fig9}
\end{figure}

In a second iteration, the first five Moffat parameters are fitted to smooth polynomial 
functions of $\lambda$ (see solid lines in Fig.~\ref{Fig8}). The position-structure parameters 
for each 
monochromatic frame are then evaluated through these polynomial functions, leaving only the 
uncalibrated flux and background as free parameters. Once we know the uncalibrated spectrum 
of feige34, $f_{\rm{S}}(\lambda)$, it is possible to build the spectral response function of 
the spectrograph $S(\lambda)$. This instrumental response is defined as the ratio between the 
detection rate in counts per second per dispersion pixel, i.e., 
$f_{\rm{S}}(\lambda)/T_{\rm{S}}$ ($T_{\rm{S}}$ is the exposure time), and the energy flux in 
erg per square centimetre per second per \AA\ at the entrance of the telescope, 
$F_{\rm{S}}^*(\lambda)$. From the extra-atmospheric energy flux of feige34 
\citep[e.g.,][]{oke90}, $F_{\rm{S}}(\lambda)$, we derive    
\begin{equation}
S(\lambda) = \frac{f_{\rm{S}}(\lambda)}{F_{\rm{S}}(\lambda)T_{\rm{S}}10^{-0.4 X_{\rm{S}} 
E(\lambda)}} ,
\label{Sfunc}
\end{equation} 
where $X_{\rm{S}}$ is the mean airmass during the observation of the calibration star, and 
$E(\lambda)$ is the standard atmospheric extinction curve at the Roque de los Muchachos 
Observatory \citep{king85}. Fig.~\ref{Fig9} presents the instrumental response in 
Eq.~\ref{Sfunc}. The sensitivities at blue wavelengths (blue line) are appreciably lower than 
those at red wavelengths (red line). A smoothed response function is also plotted in 
Fig.~\ref{Fig9} (green line). We use a 5-point filter in both arms. The red part of $S(\lambda)$ 
contains several 
noteworthy features. Apart from fringe effects, there is a striking loss of spectral 
sensitivity near 6900, 7300, 7600 and 8800 \AA. We remark that the standard extinction curve
\citep{king85} does not account for telluric absorption by molecular oxygen and water vapour.
This produces, e.g., the telluric oxygen artefacts near 6900 and 7600 \AA. In order to 
correct the molecular absorption bands, one must divide Eq.~\ref{Sfunc} by a molecular 
absorption curve $\epsilon_{\rm{S}}(\lambda)$. 

\subsection{Results}
\label{sec:4.3}

\begin{figure}
\includegraphics[width=\linewidth]{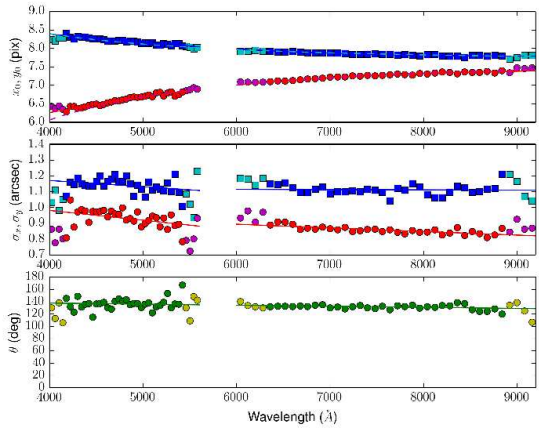}
\caption{PSF fitting results for 40 (blue arm) + 40 (red arm) spectral slices of q0957 data. 
The five parameters are also fitted to smooth polynomial functions, excluding the spectral 
edges in magenta, cyan and yellow (solid lines). Top: $x_{\rm{A}}$ (blue and cyan squares) and 
$y_{\rm{A}}$ (red and magenta circles). The dashed lines describe the wavelength-dependent 
displacements of the centroid of the A image arising from differential atmospheric refraction  
\citep{filippenko82}. Middle: $\sigma_x$ (blue and cyan squares) and $\sigma_y$ (red and 
magenta circles). Bottom: $\theta$ (green and yellow circles).}
\label{Fig10}
\end{figure}

The key ideas to extract the final q0957 spectra are outlined in Sect.~\ref{sec:4.1}. The lens 
system model consists of two $\beta$ = 3 Moffat distributions, as well as a de Vaucouleurs 
profile convolved with a $\beta$ = 3 Moffat PSF. To avoid boundary biases, this last 
numerical convolution is calculated in a square area nine times larger than the field of 
view. In a first iteration, 
we fit the centroid of the A image ($x_{\rm{A}}$, $y_{\rm{A}}$), the PSF parameters 
($\sigma_x$, $\sigma_y$, $\theta$), the orientation of the frame, the total fluxes of the two 
quasar images and the lensing galaxy ($f_{\rm{A}}$, $f_{\rm{B}}$, $f_{\rm{G}}$), and the sky 
level, to 2D data for different wavelength slices. The data cubes are split into 40 slices 
along the spectral axis, as done with the stellar data in Sect.~\ref{sec:4.2}, and the first five 
parameters are then treated as polynomial functions of $\lambda$ (see Fig.~\ref{Fig10}). To 
trace the polynomial laws, we exclude the values in the spectral edges. These excluded values 
are drawn in magenta, cyan and yellow in Fig.~\ref{Fig10}. The orientation of the frame is 
also set to the median value for all wavelength slices. In the top panel of Fig.~\ref{Fig10}, 
$y_{\rm{A}}$ is shown with red and magenta circles. Differential atmospheric refraction (dashed 
lines) causes vertical displacements $\delta y_{\rm{A}}$ exceeding one spatial pixel 
(= 0\farcs83; see Fig.~\ref{Fig6}). 

\begin{figure}
\includegraphics[width=\linewidth]{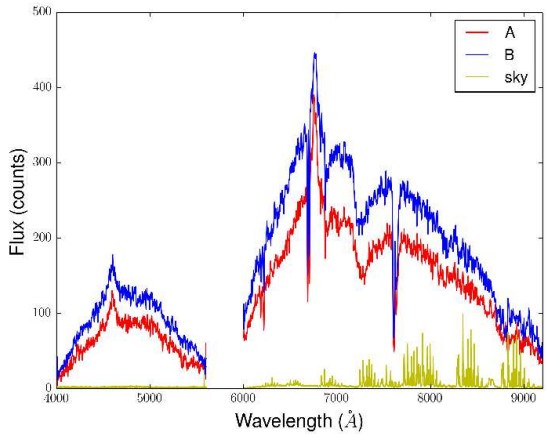}
\caption{Flux-uncalibrated spectra of the two quasar images A and B. An 8 \AA\ filter is used 
to smooth the original data. We also display the sky spectrum (counts per spaxel; yellow 
line) for comparison purposes.}
\label{Fig11}
\end{figure}

\begin{figure}
\includegraphics[width=\linewidth]{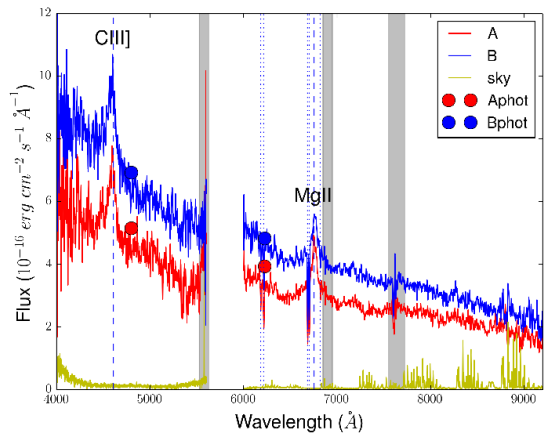}
\caption{Calibrated spectra of the two quasar images A and B. Both spectra are smoothed with 
an 8 \AA\ filter (red and blue lines). Energy fluxes from a parallel $g'r'$-band monitoring are 
also shown (red and blue circles). Vertical dashed and dotted lines correspond to emission 
and absorption lines at $z \sim$ 1.4, respectively, whereas the three gray highlighted 
regions are associated with atmospheric artefacts. Apart from two O$_2$ (telluric) absorption 
bands, an imperfect sky extraction likely produces the artefact at 5577 \AA. This wavelength 
corresponds to a strong emission line of atmospheric O\,{\sc i}. See main text for details.}
\label{Fig12}
\end{figure}

In a second iteration, we only fit $f_{\rm{A}}$, $f_{\rm{B}}$, $f_{\rm{G}}$, and the sky 
level to each monochromatic frame. The flux-uncalibrated spectra of A and B appear in 
Fig.~\ref{Fig11}. Both quasar spectra are smoothed with an 8 \AA\ filter. The galaxy spectrum 
is very noisy and is not shown in Fig.~\ref{Fig11}. However, averaging over independent 
intervals of 400 \AA, red-arm fluxes of the galaxy are comparable to quasar fluxes. This 
result agrees with previous observations of the lens system 
\citep[e.g.,][]{bernstein97,keeton98}. The response function $S(\lambda)$ in Eq.~\ref{Sfunc} 
and Fig.~\ref{Fig9} allows us to calibrate the quasar spectra (Q = A, B):  
\begin{equation}
F_{\rm{Q}}(\lambda) = \frac{f_{\rm{Q}}(\lambda)}{S(\lambda)T_{\rm{Q}}10^{-0.4 X_{\rm{Q}} 
E(\lambda)}[\epsilon_{\rm{Q}}(\lambda)/\epsilon_{\rm{S}}(\lambda)]} ,
\end{equation} 
where $F_{\rm{Q}}(\lambda)$ is the extra-atmospheric energy flux, $T_{\rm{Q}}$ and 
$X_{\rm{Q}}$ are the exposure time and airmass for q0957, and $\epsilon_{\rm{Q}}(\lambda) /
\epsilon_{\rm{S}}(\lambda)$ is the quasar-to-star molecular absorption ratio. It is assumed
that this last ratio is equal to 1, so we must be careful with possible spectral features at 
typical wavelengths for molecular (telluric) absorption bands. For example, two gray 
highlighted regions in Fig.~\ref{Fig12} correspond to telluric oxygen bands around 6900 and 
7600 \AA. The calibrated spectra of both quasar images are plotted in Fig.~\ref{Fig12} with 
red (A image) and blue (B image) lines. This figure also includes independent energy fluxes 
from our photometric monitoring with RATCam in the Sloan $g'$ and $r'$ bands \citep{sha08}. 
The red (A image) and blue (B image) circles represent the $g'r'$-Sloan fluxes, which should 
be compared with spectral fluxes averaged over the $g'$ and $r'$ broad passbands around 4800 
and 6225 \AA\ (see filter responses in Fig.~\ref{Fig12}). We can not properly asses the 
calibration in the red arm because our spectra do not cover the full $r'$ passband. However, 
the relative differences in the blue arm do not exceed 7\%. 

The most prominent features of the quasar spectra in Fig.~\ref{Fig12} are the C\,{\sc iii}] 
($\lambda$1909) and Mg\,{\sc ii} ($\lambda$2798) emission lines (see the two vertical dashed 
lines). These are observed around 4600 and 6740 \AA\ \citep[$z_{\rm{Q}}$ = 1.41;][]{walsh79}. 
Going into details, the more accurate and recent value $z_{\rm{Q}}$ = 1.414 
\citep[e.g.,][]{rao00} is very close to our redshifts from the Mg\,{\sc ii} emission line: 
$z_{\rm{Mg\,{\sc II}}}$(A) = 1.415 and $z_{\rm{Mg\,{\sc II}}}$(B) = 1.416. Additionally, it 
is well known the existence of a damped Ly$\alpha$ (DLA) system at $z_{\rm{DLA}}$ = 1.391 
\citep[e.g.,][]{rao00}, and redshifts from the Fe\,{\sc ii}+Mg\,{\sc ii}+Mg\,{\sc i} 
absorption-line complex in our spectra (vertical dotted lines in Fig.~\ref{Fig12}) deviate 
only $\delta z$ = 0.001 from $z_{\rm{DLA}}$. These results prove the high precision of the 
wavelength calibration. In regards to the signal-to-noise ratio (SNR), we obtain SNR $\sim$
10$-$20 per spectral pixel in the 6300$-$7300 \AA\ continuum. Therefore, averaging over 80 
\AA\ intervals (50 pixels), it is achieved $\sim$ 1\% accuracy in the red continuum flux. An 
estimation of the accuracy in the Mg\,{\sc ii} emission-line flux is also possible. We obtain 
a few percent error for this feature.  
      
\section{Conclusions}
\label{sec:5}

FRODOSpec data of \object{Q0957+561} for a given observation session (2011 March 1) show 
that the standard processing pipeline L2 \citep{barnsley12b} does not yield satisfactory 
results. We then introduce the new processing scheme L2LENS. This is well suited to the 
production of spectra of blended faint sources in crowded fields. All L2LENS reduction tasks 
for the FRODOSpec data on 2011 March 1 are fully detailed in Appendix A. In addition to the 
user's guide in Appendix A, our website$^1$ includes all required data files and Python 
scripts, as well as several auxiliary files. 

Firstly, L2LENS accurately extracts the raw spectrum for 
each fibre across the field of view. The long exposure of the relatively faint system 
\object{Q0957+561} is affected by a large number of cosmic rays and significant amounts of 
scattered light. While the standard pipeline does not account for these artefacts, cleaning 
of cosmic-ray events and scattered-light backgrounds (and some additional refinements) are 
incorporated into L2LENS. Secondly, the new processing software allows us to perform PSF 
fitting photometry on pixels (or broader slices) along the spectral axis, and thus, to infer 
spectra for the three blended sources of the lens system, i.e., the two quasar images and the 
lensing galaxy. Our PSF fitting method considers differential atmospheric refraction effects, 
and it is a variant of the approach by \citet{wisotzki03}. The final products of L2LENS are 
the wavelength- and flux-calibrated spectra of both quasar images. Unfortunately, L2LENS does
not produce accurate spectrophotometric data for the faint galaxy, whose light is distributed 
throughout most the field of view. Due to this extended light distribution with an effective 
radius of 4\farcs63, the spatial pixels contain a relatively weak signal from the galaxy. To 
asses the true limits of FRODOSpec, we are currently observing more compact lens systems 
including fainter quasar images.  

The new quasar spectra contain emission and absorption lines whose redshifts deviate $\leq$ 
0.002 from the expected ones. Moreover, it is achieved a few percent accuracy in red 
continuum and Mg\,{\sc ii} emission-line fluxes. This demonstrates the big potential of 
robotic programs with FRODOSpec. We also obtain 1$-$10\% uncertainty in image flux ratios in 
the region of 6300$-$7300 \AA, which proves that FRODOSpec is competitive with the Space 
Telescope Imaging Spectrograph \citep{goico05}. We hope that our non-standard FRODOSpec data 
reduction will stimulate other teams to conduct spectroscopic projects involving relatively 
faint and/or blended sources (long exposures and/or crowded fields).

Future, relatively large optical telescopes with integral field spectrographs and flexible 
schedulings will be extraordinarily well suited for surveys/monitorings of blended faint 
sources with $V \geq$ 20 mag. For example, the successor to the Liverpool Robotic 
Telescope\footnote{http://telescope.livjm.ac.uk/lt2/} might be equipped with an improved
version of the FRODOSpec spectrograph. This hypothetical integral field unit should have a 
larger field of view ($\sim$ 1\arcmin\ on sky) and smaller spatial pixels ($\sim$ 0\farcs1 on 
sky). The spatial improvements are critical to extract empirical PSFs from field stars and 
accurately resolve all sources in very crowded fields. 

\acknowledgements
We would like to thank Robert Barnsley and the anonymous referee for several interesting 
comments and suggestions. The Liverpool 
Telescope is operated on the island of La Palma by Liverpool John Moores 
University in the Spanish Observatorio del Roque de los Muchachos of the Instituto de 
Astrof\'{\i}sica de Canarias with financial support from the UK Science and Technology 
Facilities Council. We thank the Liverpool Telescope staff for kind interaction over the 
observation period. The Liverpool Quasar Lens Monitoring program (XCL04BL2) is supported by 
the Spanish Department of Science and Innovation grant AYA2010-21741-C03-03 (GLENDAMA 
project), and the University of Cantabria.

\newpage

\appendix

\section{L2LENS}
This simple User's Guide describes all steps to successfully process the FRODOSpec 
observations of \object{Q0957+561} (q0957) and the calibration star \object{Feige 34} 
(feige34) on 2011 March 1. 
We assume that /home/user/l2lens is the path to the directory (folder) \verb|l2lens|. 
The \verb|l2lens| folder houses the Python scripts of the L2LENS software, some auxiliary 
files and the subfolder \verb|110301|. All relevant data files of q0957, feige34, and the W 
and Xe lamps are located in \verb|110301|. There is also a sub-subfolder \verb|database| 
containing additional auxiliary files. The initial \verb|l2lens| folder and a README file 
(explaining its contents; see also here below for a description of the main files and its 
usage) are freely available at the GLENDAMA website$^1$. This online material is 
distributed as a single compressed file l2lens.zip. We also assume that IRAF and Python are 
properly installed. The Python modules PyRAF, PyFITS, NumPy, SciPy and Matplotlib are needed 
to run L2LENS scripts. 

\subsection{Fibre spectrum extraction (IRAF V2.16 + Python 2.7)}
For convenience, the pre-processing pipeline (L1) outputs for the blue arm of FRODOSpec are 
renamed with shorter labels. This is done in \verb|110301|, where the data files are found,
using the IRAF command lines:
\begin{verbatim}
cl> imcopy b_w_20110301_2_1_1_1 bw          	
cl> imcopy b_e_20110301_7_1_1_2.fits[0] be1 	
cl> imcopy b_a_20110301_8_1_1_1 ba1         	  
cl> imcopy b_e_20110301_9_1_1_2.fits[0] be2 	
cl> imcopy b_a_20110301_10_1_1_1 ba2        
\end{verbatim}
Similar FITS files rw, re1, ra1, re2 and ra2 are produced for the red arm. Here, *w, *e1, 
*a1, *e2 and *a2 refer to W lamp, q0957, Xe lamp for q0957, feige34 and Xe lamp for feige34,
respectively.

The next steps are: 
\begin{enumerate}
  \item {\it Removing cosmic-ray events}\\
Based on the spectroscopic version of the L.A.Cosmic 
algorithm$^4$. This is put 
into the \verb|extern| directory of IRAF (/iraf/iraf/extern), and then  
\begin{verbatim}
cl> task lacos_spec = /iraf/iraf/extern/ 
lacos_spec.cl 
cl> stsdas
cl> lacos_spec be1 be1cr be1m.pl 
gain=2.134 readn=3.85 	  
cl> lacos_spec re1 re1cr re1m.pl 
gain=2.350 readn=4.44
\end{verbatim}
A similar procedure is followed for cleaning the feige34 frames be2 and re2. After 
subtraction of cosmic-ray events, the main FITS files of q0957 and feige34 are be1cr, re1cr, 
be2cr and re2cr\\
  \item {\it Finding and tracing fibre positions along the dispersion axis}\\
Based on the IRAF/SPECRED package. The command lines are 
\begin{verbatim}
cl> noao
cl> imred
cl> specred
cl> apall bw nfind=144 resize- lower=-2 
upper=2 background- minsep=5 maxsep=10 
width=7 weights- clean- t_func="legendre" 
t_step=50 t_niter=1 t_order=3 t_sample=
"500:2500"
cl> apall rw nfind=144 t_order=6 
t_sample="1450:3450"
\end{verbatim}
The FITS files bw.ms and rw.ms contain the blue-arm and red-arm W-lamp (continuum emission) 
spectrum for each fibre\\
  \item {\it Removing scattered-light backgrounds}\\
Based on the IRAF/SPECRED package  
\begin{verbatim}
cl> apscatter be1cr be1sc ref=bw 
buffer=0 apscat1.order=5 apscat2.order=5 
apscat2.sample="5:4096" inter- 
\end{verbatim}
A similar procedure is followed for cleaning the q0957 frame in the red arm, i.e., re1cr 
$\to$ re1sc. Now the main FITS files of q0957 are be1sc and re1sc\\
  \item {\it Fibre flux extraction}\\
Based on the IRAF/SPECRED package  
\begin{verbatim}
cl> apall be1sc ref=bw out=be1.ms trace- 
recen- intera-
cl> apall ba1 ref=bw trace- recen- 
intera-
cl> apall be2cr ref=bw out=be2.ms trace- 
recen- intera-
cl> apall ba2 ref=bw trace- recen- 
intera-
\end{verbatim}
The FITS file be1.ms (ba1.ms) contains the q0957 (Xe-lamp) spectrum in the blue arm for each 
fibre, while the file be2.ms (ba2.ms) includes the blue-arm feige34 (Xe-lamp) spectra. Using 
similar commands, it is also possible to extract spectra in the red arm\\ 
  \item {\it Wavelength calibration (dispersion solution)}\\
Based on the Python script reident.py 
\begin{verbatim}
./reident.py
\end{verbatim}
This program uses the idba0.ms and idra0.ms files (approximated solutions) in the 
\verb|database| sub-subfolder. The lists of Xe emission lines are frodo\_blue.dat (blue arm) 
and frodo\_red.dat (red arm). Both lists are available in \verb|l2lens|. It is necessary to 
run the Python script four times: \verb|name = 'ba1'|, \verb|'ba2'|, \verb|'ra1'| and 
\verb|'ra2'| in reident.py\\  
  \item {\it Throughput correction}\\
Based on the Python script norm.py 
\begin{verbatim}
./norm.py
\end{verbatim}
Run this script twice: \verb|arm = 'b'| and \verb|'r'| in norm.py. The normalized spectrum 
for each fibre can be found in the FITS files be1nr.ms (q0957/blue arm), re1nr.ms (q0957/red 
arm), be2nr.ms (feige34/blue arm) and re2nr.ms (feige34/red arm)\\
  \item {\it Spectral rebinning (dispersion correction)}\\
Based on the Python script disp\_cor.py 
\begin{verbatim}
./disp_cor.py 
\end{verbatim}
Run the script twice: \verb|arm = 'b'| and \verb|'r'| in disp\_cor.py. This gives the final 
raw spectrum for each fibre: be1dc.ms (q0957/blue arm), re1dc.ms (q0957/red arm), be2dc.ms 
(feige34/blue arm) and re2dc.ms (feige34/red arm)\\
  \item {\it Making spectral data cubes}\\
Each data cube gives the 2D flux in the 12$\times$12 fibre array at each wavelength pixel. We 
use the script rss\_cube.py
\begin{verbatim}
./rss_cube.py 
\end{verbatim}
Run the script four times: \verb|inname = 'be1'|, \verb|'be2'|, \verb|'re1'| and 
\verb|'re2'| in rss\_cube.py. This produces the data cubes (FITS files) b1 (q0957/blue arm), 
r1 (q0957/red arm), b2 (feige34/blue arm) and r2 (feige34/red arm)
\end{enumerate}

\subsection{Flux-calibrated spectra of the lens system (Python 2.7)}
\begin{enumerate}
  \item {\it Feige 34: photometry on slices along the spectral axis}\\
Based on the Python script m2free.py. Both spectral data cubes (b2 and r2) are split into 40 
slices along the spectral axis. Each slice is fitted to a seven-parameter model, where the 
free parameters are: the centroid of the star, the FWHM along major and minor axis, the 
position angle of the major axis, the uncalibrated flux of the star, and the background sky 
level. We use the command line \verb|./m2free.py directory x0 y0|, where \verb|directory| 
refers to the subfolder containing the two data cubes, and \verb|x0 y0| is an approximate 
stellar centroid, i.e., 
\begin{verbatim}
./m2free.py 110301 6 3 
\end{verbatim}
The b2.free and r2.free outputs show 40 solutions each (one per slice)\\
  \item {\it Feige 34: polynomial fits to position-structure parameters}\\
Based on the Python script m2fit.py. The five position-structure parameters are fitted to 
smooth polynomial functions of wavelength 
\begin{verbatim}
./m2fit.py 110301 
\end{verbatim}
The b2.fit and r2.fit outputs show 2001 solutions each (one per dispersion pixel). These fits 
are displayed in Fig. 8\\
  \item {\it Feige 34: photometry on monochromatic frames}\\
Based on the Python script m2fix.py. Each monochromatic frame is fitted to a two-parameter 
model. The five position-structure parameters are evaluated through smooth polynomial 
functions (b2.fit and r2.fit), leaving only the uncalibrated flux and the sky level as free 
parameters  
\begin{verbatim}
./m2fix.py 110301 
\end{verbatim}
The b2.fix and r2.fix outputs show 2001 (stellar flux, sky level) pairs each. These files 
also include the standard spectral response of FRODOSpec (Fig. 9), which is based on the true 
spectrum of Feige 34 (f34a.oke in \verb|l2lens|) and the standard atmospheric extinction 
curve at the Roque de los Muchachos Observatory (lam\_extin.dat in \verb|l2lens|). The 
optional script m2graph.py (use the command line \verb|./m2graph.py 110301|) also allows the 
user to check b2.fix and r2.fix\\ 
  \item {\it Lens system: photometry on slices along the spectral axis}\\
Based on the Python script m1free.py. Both spectral data cubes (b1 and r1) are split into 40 
slices along the spectral axis. Each slice is fitted to a ten-parameter model, where the free 
parameters are: the centroid of Q0957+561A, the FWHM along major and minor axis, the position 
angle of the major axis, the orientation of the frame, the uncalibrated total fluxes of the 
two quasar images (Q0957+561A and Q0957+561B) and the lensing galaxy, and the sky level. We 
use the command line \verb|./m1free.py directory x0 y0|, where \verb|directory| refers to the 
subfolder containing the two data cubes, and \verb|x0 y0| is an approximate centroid of 
Q0957+561A, i.e., 
\begin{verbatim}
./m1free.py 110301 8 6.5  
\end{verbatim}
The b1.free and r1.free outputs show 40 solutions each (one per slice)\\
  \item {\it Lens system: polynomial fits to position-structure parameters and estimation of 
orientations}\\
Based on the Python script m1fit.py. The five position-structure parameters are fitted to 
smooth polynomial functions of wavelength, excluding the spectral edges. For each spectral 
arm, it is also obtained the median orientation for all wavelength slices
\begin{verbatim}
./m1fit.py 110301  
\end{verbatim}
The b1.fit and r1.fit outputs show 2001 solutions each (one per dispersion pixel). These fits 
are displayed in Fig. 10. Apart from position-structure data, the outputs contain the 
orientations of each arm\\
  \item {\it Lens system: photometry on monochromatic frames}\\
Based on the Python script m1fix.py. Each monochromatic frame is fitted to a four-parameter 
model. The five position-structure data and the frame orientation are taken from the b1.fit 
and r1.fit files. Thus, only the uncalibrated fluxes and the sky level are free parameters  
\begin{verbatim}
./m1fix.py 110301  
\end{verbatim}
The b1.fix and r1.fix outputs include 2001 (flux\_A, flux\_B, flux\_G, sky level) vectors 
each\\
  \item {\it Final spectra of the two quasar images}\\
Based on the Python script m1graph.py  
\begin{verbatim}
./m1graph.py  
\end{verbatim}
The b.dat and r.dat files contain the flux-calibrated spectra of the two quasar images and 
the lens galaxy (units are described in the main text). The m1graph.py script also produces 
Fig. 11 and a variant of Fig. 12 
\end{enumerate}

\end{document}